# Long-Lived Highly Emissive MOFs as Potential Candidates for Multiphotonic Applications

Mario Gutiérrez,[a] Cristina Martín,*[b,c] Johan Hofkens,[b] and Jin-Chong Tan*[a]

Long-lived emissive materials based on room temperature phosphorescence (RTP) and thermally activated delayed fluorescence (TADF) are considered as the cornerstone of the development of optical sensors, security systems and solid-state lighting. Nevertheless, molecular systems with these properties are scarce because most of them suffer from aggregation caused quenching emission (ACQ). One approach to address this shortcoming is by inhibiting the molecular motions/vibrations by employing a fixed matrix as afforded by a metal-organic framework (MOF). There, the organic chromophores are confined in a crystalline framework, and the structure-property relationship can be designed to get RTP/TADF. Inspired by this, the present work explores the relation between the linker arrangement and the physicochemical properties of two isochemical MOFs with different crystalline structures. The denser MOF exhibits a long-lived green RTP due to a hyperfine coupling of the linkers. On the other hand, the more porous MOF presents a long-lived temperature-dependent turquoise emission, reflecting the influence of the TADF. Hence, this study provides a huge advance about the potential of MOFs to undergo RTP and TADF emission, and at the same time, demonstrates their potential applicability in a wide range of photonic technologies, including physical and chemical sensing and the first example of a MOF-LED based on RTP-MOFs.

## Introduction

Over the last years, photonics, the science for harnessing light, has been pinpointed as a key enabling technology for sustainable world-wide growth through the provision of practical and cost-effective solutions.[1] Inside this broad field, new luminescent materials (luminogens) with multifunctional applications are considered as essential to unlock the potential of photonic technologies such as sensing, solid-state lighting and communications.[2, 3] Since the performance of these applications are heavily dependent on the material's physicochemical properties, the intermolecular interactions, which normally weaken their emission by aggregation caused quenching (ACQ),[4-6] must be understood and controlled to develop highly emissive materials in aggregated state. Indeed, recent reports have shown that certain intermolecular interactions may preclude intramolecular motions, suppressing the non-radiative deactivation pathways, and enhancing the emission properties.[6-8] Thus, a proper control of the location of the fluorophores is paramount to improve the luminescence properties of solid materials.

On this regard, Luminescent Metal-Organic Frameworks (LMOFs), a class of materials with multiple photonic building blocks (organic linkers, metal ions or guests), connected in a modular and ordered structure offer a perfect alternative to address the abovementioned limitations.[9-11] Knowing the position of each component within the periodic framework allows one to unveil all the interactions and the synergies between the building blocks. Although a vast number of possible synthetic combinations have led to numerous new LMOFs,[12-14] the use of these crystal platforms for suppressing the molecular motions and the non-radiative recombination by enhancing the room temperature phosphorescence (RTP) is rarely explored. The strategy to build this type of LMOFs, known as RTP-MOFs, is thus far focused on the nucleation and growth of the MOF.[15] Therefore, factors like organic linker or metal centres, the cornerstone for MOF functionality up until now, are becoming secondary in favour of other parameters like solvent, which is directly responsible for determining the MOF topology and structure.[16-19]

To date, the effect of solvent on MOF formation and the subsequent modification of their optoelectronic properties are rarely studied, and what is even less understood is the effect on generating defects in MOFs during synthesis. Considering this, we present a systematic study on how the 3D MOF (Pb-BDC MOF) structure is modified by using two types of solvent during the synthesis and how this leads to different underlying photophysical mechanisms. Whereas the main mechanism of the MOF synthesised in water is RTP ($\sim$ 2 ms), which makes this material an excellent candidate for optical thermometer and lighting applications, the use of dimethylformamide generates a more porous structure, inhibiting partially the RTP while favouring a TADF mechanism. Furthermore, we show the solvent affects the presence of defects in the MOF structure, thereby highlighting the importance for understanding their existence on optoelectronic applications. This study stresses the primary role of the solvent for controlling the interaction between the basic building blocks, as it can template different MOF structures based on the linker packing.

## Results and Discussion

The synthesis and characterization (structural, morphological, chemical, and thermogravimetric) of Pb-BDC MOFs are described in the Supporting Information. Briefly, both MOFs were synthesised by following the same procedure, where just the solvent of the reaction was modified. The solvent will induce a different crystallization, leading to a more densely packed structure when water is used (Pb-BDC/$H_2O$), or resulting in a more porous arrangement in the MOF when mediated by *N*, *N*-dimethylformamide (Pb-BDC/DMF). Interestingly, these structural changes have a strong impact on the luminescent properties and their possible applicability, even though both MOFs have almost the same chemical composition.

### Photophysical Studies

To begin with, the steady-state emission properties of these Pb-BDC MOFs were explored and found to be completely different from one each other. The Pb-BDC/$H_2O$ MOF presents a green emission with its intensity maximum at 525 nm, and an outstanding high emission quantum yield (QY) of 59% in powder form (Figure 1A). On the other hand, Pb-BDC/DMF MOF exhibits a turquoise emission with its maximum located at 480 nm and a lower (but still excellent) QY of 32% in solid state (Figure 1B). These differences may be ascribed to the distinct packing of the BDC organic linkers in those two framework structures. It is well known that terephthalic acid (BDC) is almost non emissive in solution form, however the crystallization of these types of aromatic compounds can induce a room temperature phosphorescence (RTP) through the formation of a radical ion pair (RIP) with the help of hyperfine coupling, or thermally activated delayed fluorescence emission (TADF).[20-22] The emission through RTP or TADF will strongly depend on the packing of the linkers.[23] Therefore, the way of crystallization within the MOF will strongly affect their luminescence properties. As expected, the emission of the more densely packed structure (Pb-BDC/$H_2O$), where the interlinker distance and orientation is more favorable, will be ruled by the RTP mechanism. However, in the more porous structure, the larger interlinker distances will trigger a competition between the RTP and TADF mechanisms. To shed more light on these photophysical phenomena, the time-resolved emission properties of both MOFs at the microsecond time scale regime were explored.

The long-lived emission lifetime of Pb-BDC/$H_2O$ and Pb-BDC/DMF were collected at 77 K and 298 K. As shown in Figure 2, the Pb-BDC/$H_2O$ MOF decays multiexponentially with relaxation times of $\tau_1$= 39 ns, $\tau_2$= 145 μs and $\tau_3$= 1.53 ms (Figure 2a and Figure S5) independently of the recorded temperature, while the emission decays of Pb-BDC/DMF depend on the temperature, with time components of $\tau_1$= 43 ns, $\tau_2$= 0.29 μs and $\tau_3$= 15 μs at 298 K, and a lack of long emission component at 77 K (Figure 2b and Figure S5). Besides this, the emission lifetime of Pb-BDC/$H_2O$ is much longer than that of Pb-BDC/DMF, supporting the fact that the more packed structure will trigger the emission by the RTP mechanism. Additionally, the time-resolved emission spectra (TRES, Figure 2c, bottom) of Pb-BDC/$H_2O$ match well with that recorded under steady-state, reinforcing the RTP mechanism. The TRES spectra of Pb-BDC/$H_2O$ was also recorded at 77 K (Figure S6), showing no significant changes to that collected at 298 K, reflecting that this mechanism does not depend on the temperature. On the other hand, the long lifetime observed for Pb-BDC/DMF at 298 K also reflects an emission from a forbidden state, however the TRES (Figure 2c, top), which is shifted to lower wavelengths, could indicate a competition between RTP and TADF, caused by a less packed configuration of the BDC linkers in the structure. These

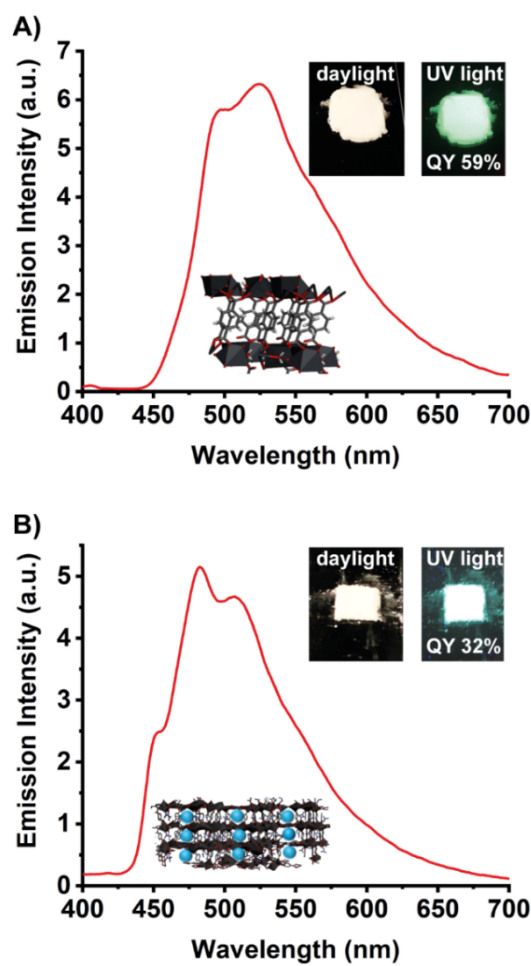

**Figure 1**. Emission spectra of **A)** Pb-BDC/$H_2O$ and **B)** Pb-BDC/DMF MOFs. The insets are a real photo of both MOFs under UV and daylight, and a representation of their 3D structure. The emission QY of both materials is also shown as inset.

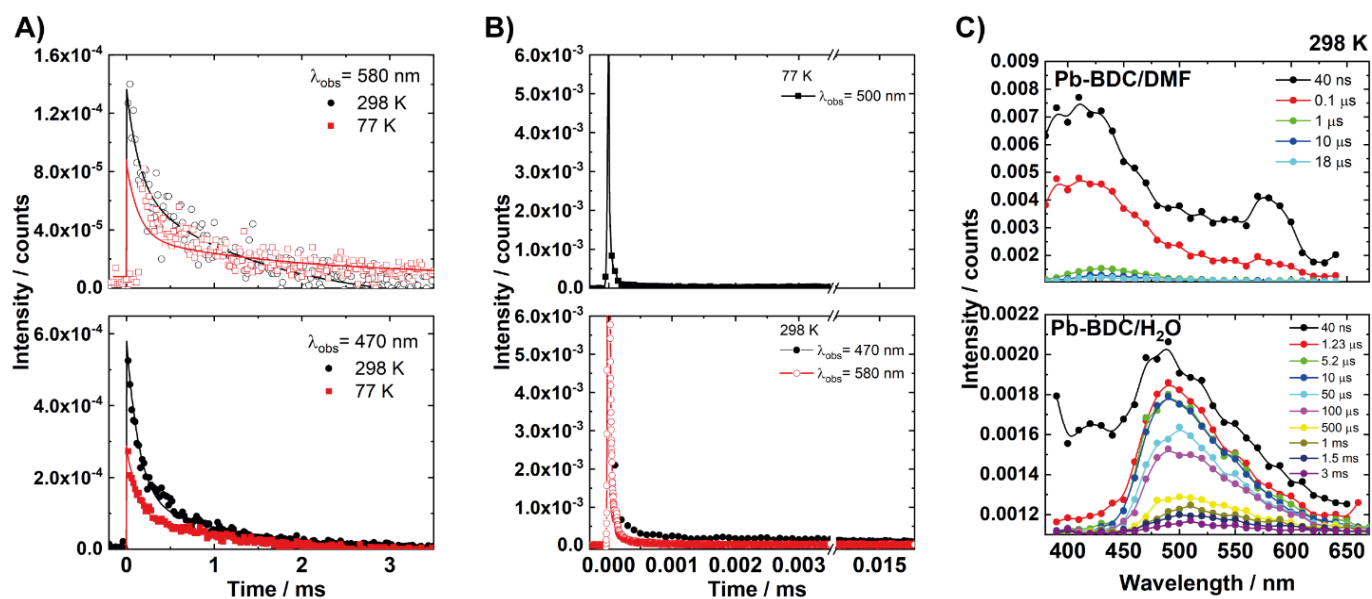

**Figure 2.** Long-lived emission decays of **A)** Pb-BDC/$H_2O$ and **B)** Pb-BDC/DMF MOFs recorded at 298 and 77 K. **C)** Time-resolved emission spectra of Pb-BDC/DMF (top) and Pb-BDC/$H_2O$ (bottom) recorded at 298 K.

results agree with previous reported systems, where the molecular packing controls the TADF/RTP mechanisms.[23-25] Furthermore, the TADF mechanism is further reinforced by the lack of a long-lived component at 77 K, while it is clearly visible at 298 K, reflecting a mechanism which is temperature-dependent as expected for a TADF phenomenon.[26] Note that the band at 400 nm appearing just at short delay times may be attributed to the prompt emission of BDC linkers, as it has been previously reported for other BDC-based MOFs.[27, 28]

**Applications**

Encouraged by the high emission QY and the RTP/TADF properties of the Pb-BDC MOFs, we have delved into a wide range of possible photonic applications where both phenomena play a key role.

**Chemical Detection by Luminescent Solvatochromism**

The photoresponse of the RTP/TADF-LMOFs to the presence of different volatile organic compounds (VOCs) was explored by immersing 4 mg of each MOF in 4 mL of the corresponding solvent (see SI). The luminescence of Pb-BDC/$H_2O$ is strongly affected by the presence of different VOCs, showing spectral shifts and changes in its emission intensity and in the CIE coordinates (Figure S7A-C). This can be caused by the presence of linker defects created during the synthesis. It is well known that protic solvents, like water, can modulate the self-assembly reaction between the metals during the synthesis because the presence of higher amounts of protons might induce a hydrolysis of the reaction, which finally culminates in a linker vacancies formation.[18, 19, 29-31] Therefore, the solvents could interact with the open metal sites, creating changes in the structure, which will modify the interlinker distance and angle, affecting the hyperfine coupling, and subsequently the RTP phenomenon. This effect is evident when Pb-BDC/$H_2O$ is in presence of water, as in this case, instead of maintaining the photonic properties of the MOF in the solid state, an additional band at 400 nm appears, which indicates the fluorescence emission of the linker itself (Figure S7A).[23]

On the other hand, the emission of Pb-BDC/DMF in the different solvents is less altered in intensity still the final color is slightly modified as shown by the CIE coordinates (Figure S7F). Surprisingly, the role of solvent is more noticeable in the case of water and acetone (Figure S7D-F). When immersing the Pb-BDC/DMF MOF in water, its emission spectrum completely matches that observed for Pb-BDC/$H_2O$, indicating a MOF-to-MOF structural transformation. To prove this assumption, the Pb-BDC/DMF MOF was soaked in water for 5 h, then dried, and the obtained powder was characterized by PXRD, which clearly shows the same diffraction peaks to those observed for Pb-BDC/$H_2O$ (Figure S9), reflecting a solvent-induced MOF-to-MOF transformation. Additionally, and very remarkably, Pb-BDC/DMF is very sensitive to the presence of acetone as its emission is completely quenched (Figure S7D-E). Indeed, a suspension of this MOF in DMF solvent is able to detect low volumes of acetone (Figure S8), indicating a high selectivity towards this analyte. Even though more tests are required, these preliminary results point out that Pb-BDC/DMF MOF could be a promising material to be used in the detection of low levels of acetone to control and monitor individuals suffering from diabetes.[32-34]

**Luminescent Thermochromism**

The ability of both MOFs to detect fluctuations in the temperature through changes in their emission properties was also tested. Pb-BDC/$H_2O$ exhibits a lineal decrease of its emission intensity maintaining the same CIE coordinates upon increasing the temperature (up to 125 ºC, Figure 3). This emission quenching behaviour can probably be caused by alterations of the network with high temperatures, where the vibrations of the network may decouple the linker interaction, blocking the RTP process. Remarkably, the initial emission intensity can be recovered upon cooling the sample at room temperature (Figure 3B), and this cyclic test (heating and cooling) was reproduced 10 times without significant loses in the efficiency of the emission and sensing ability (Figure 3C). Thus, the Pb-BDC/$H_2O$ is an ideal candidate to construct a luminescent thermometer able to replace conventional thermometers in specific environments, where the presence of high electromagnetic fields precludes their use.[35, 36] Most of the LMOF-thermometers reported hitherto are based on pricey rare earth elements,[9, 37, 38] which are functional in the range of 10-350 K, while the MOFs presented here are inexpensive and work effectively in the range of ~273-400 K, spanning a wide possible range of applications.

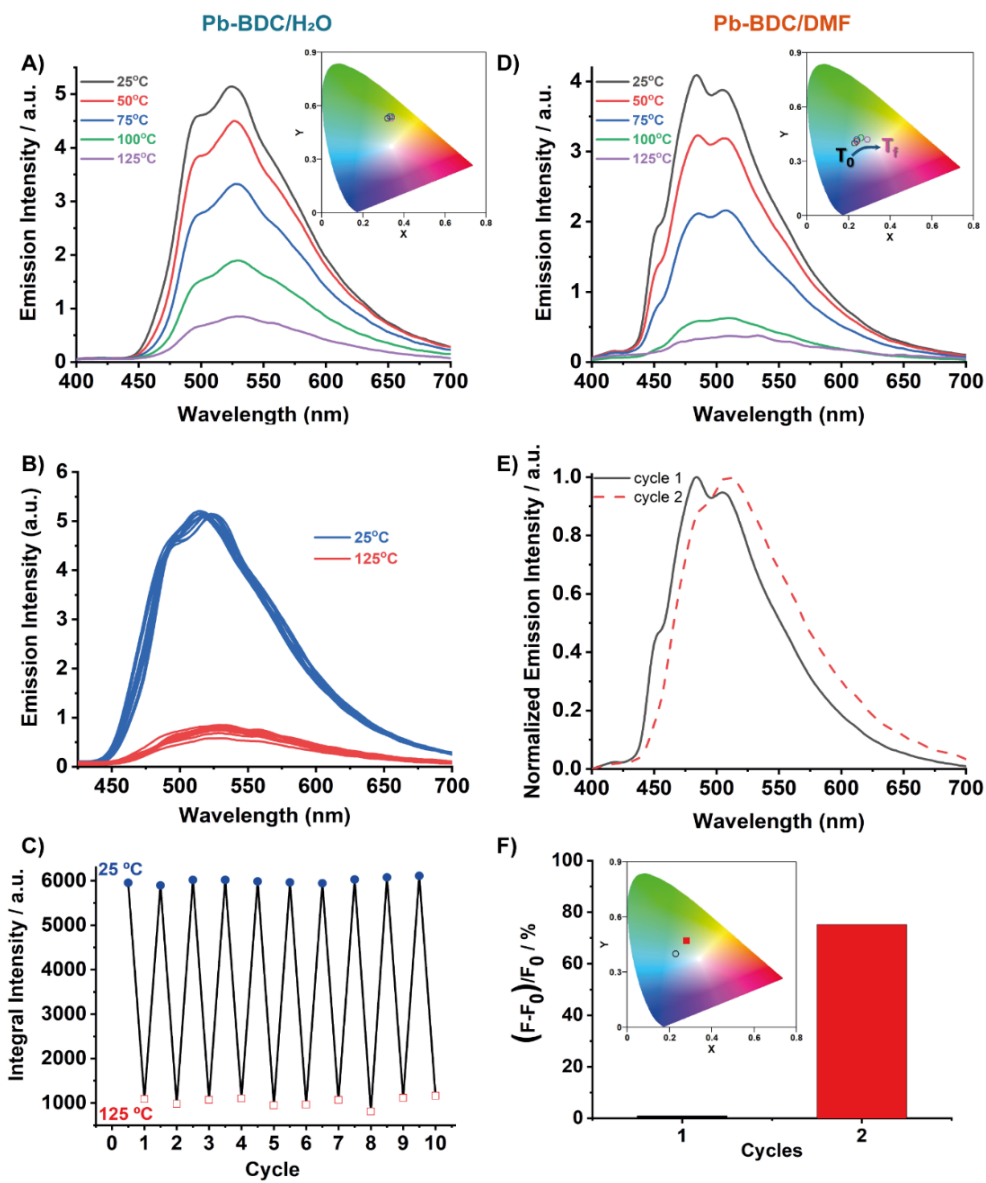

**Figure 3**. Emission spectra of **A)** Pb-BDC/$H_2O$ and **D)** Pb-BDC/DMF MOFs recorded at different temperatures. **B), E)** Emission spectra of **B)** Pb-BDC/$H_2O$ and **E)** Pb-BDC/DMF recorded during several cycles (two for Pb-BDC/DMF) of heating (125°C) and cooling (25°C) the sample. **C), F)** Representation of the emission intensity of **C)** Pb-BDC/$H_2O$ and **F)** Pb-BDC/DMF MOFs after heating and cooling the sample for several cycles (two for Pb-BDC/DMF). The inset are the CIE coordinates of the emission of the Pb-BDC MOFs at different temperatures.

On the other side, Pb-BDC/DMF is also sensitive to high temperatures, presenting a similar emission quenching and an additional red shift of the emission spectrum (Figure 3D). However, in this case, the initial emission intensity is not retrieved after cooling down the sample (Figure 3F). This can be an advantage for the fabrication of smart packing thermometers, as this material could retain the information on temperature history when exposed to high temperatures, which can be easily monitored with a UV light torch, as the color dramatically change from blue to green (inset of Figure 3C), indicating by this way a damage of the product. The shift in the emission spectrum and the irreversibility of the process is produced by a structural transformation of the MOF due to the removal of DMF molecules. At high temperatures, the DMF molecules are desorbed from the material and the crystalline structure transforms to a denser one as shown by the PXRD pattern of the Pb-BDC/DMF material heated at 125ºC (Figure S11).[39]

**Light Emitting Diodes (LEDs)**

To further explore other possible applications of these RTP/TADF-LMOFs, the Pb-BDC MOFs were implemented as electroluminescence materials in a bottom-up light emitting diode (LED). Based on our previous experience with MOF-LEDs,[40-42] where the single layer configuration did not give any stable LEDs performance, additional layers were included, improving the energy alignment between all the layers, and obtaining a better balance in the carrier mobility all at once. Therefore, the device was fabricated employing the following configuration: ITO anode/PEDOT:PSS (40 nm)/ Pb-BDC (150 nm)/ TPBi (5 nm)/ Al metal cathode (150 nm) (inset of Figure 4A). Stable working LED devices were manufactured with this protocol with turn-on voltages of ~2 V and 4 V for Pb-BDC/H$_2$O and Pb-BDC/DMF, respectively (Figure 4A). Under forward voltage in Figure 4B (top), the electroluminescence (EL) spectra consist of a broad band spanning from 400 to 700 nm which becomes more intense at the same voltage (12 V) for Pb-BDC/H$_2$O (Figure 4B, top). As it was previously observed for other MOFs,[40-42] the EL spectra differ from the photoluminescence (PL) ones leading to a change in the emission color as depicted in the CIE coordinates (Figure 4C). Although the use of these materials as electroactive layers in LED instead of using as down-light converters modify the final display color, the best light quality shown by CRI test (above 80 like the commercial ones) highlights the performance of these MOFs in electroluminescent devices.

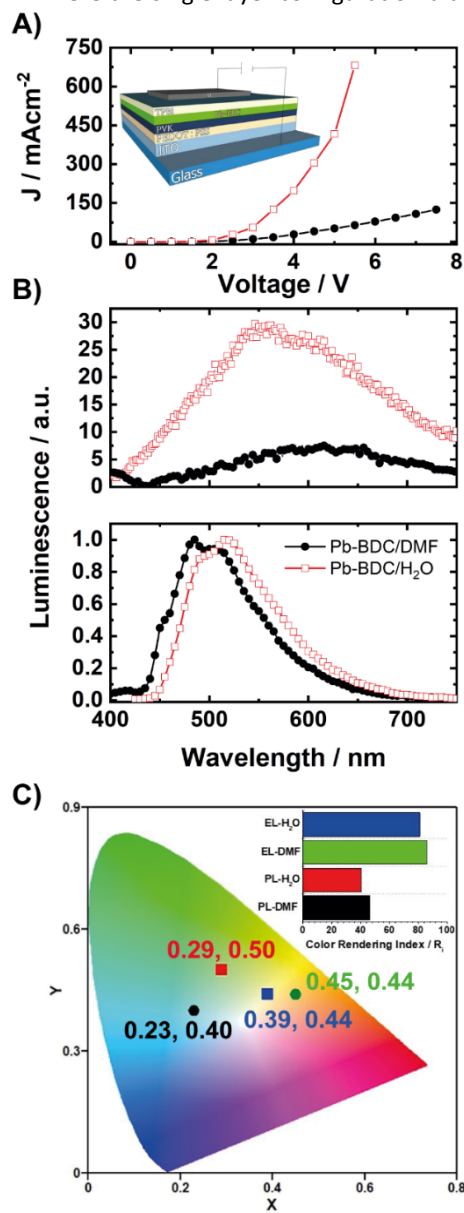

**Figure 4.** Comparison of A) the current density versus voltage curves and B) their corresponding electroluminescence spectra (top) and photoluminescence (bottom) of the Pb-BDC/DMF (black circles) and Pb-BDC/H$_2$O (red full squares). The inset in A show a schematic representation of the layers used in the devices. C) CIE coordinates of the EL at 6 V. The inset display in C shows the CRI values for each sample.

To shed light on the origin of the broader and redder EL spectra, a gaussian deconvolution was performed for the signal derived from both devices (Figure S12). In both cases, two bands are required to satisfy the fit, one centered at ~510 nm and the other one at ~630 nm. The bluest emission component matches with the maximum of the PL spectra, and therefore can be attributed to the RTP/TADF emission. On the other hand, as it was observed previously for other MOF-LEDs,[40-42] defects can originate a red emitting species by acting as deep electronic traps. Hence, these Pb-BDC MOFs which are not free of defects (*vide supra*) will be also susceptible to having electron-hole recombination in those trap states, red-shifting the EL spectra. However, the smaller relative contribution of the 630 nm component (previously assigned to defects recombination) in comparison to the other Pb-BDC/DMF (65 % versus 80 %) ruled out this possibility. A deeper characterization of both devices based on single carrier's devices pointed out that the most plausible explanation is related to MOF structure, where the denser structure of Pb-BDC/H$_2$O will benefit the emissive triplet state generation and the charge mobility recombination processes (see part 6 in SI). These results are in agreement with other reported studies,[43-47] where in those it could be observed than stronger packing not only reinforce a longer emission lifetime (RTP) but also boost the charge transfer processes. Therefore, the differences observed in the optoelectronic properties are attributed more to the MOF structure than to the presence of defect.

**Conclusions**

Herein, we demonstrated that the crystallization induced emission of MOFs is an outstanding approach for obtaining solid-state RTP and TADF emitter materials. Those materials are considered as one of the most promising materials for developing a wide range of photonic key technologies. Indeed, we showed their promising potential in optical sensing (temperature or chemicals) and in solid-state lighting applications, highlighting the importance of understanding the connection between the crystalline structure (packing rearrangement) and the photophysical mechanisms, and opening an almost unexplored pathway for the development of a new family of RTP-MOFs.

## Author Contributions

**Mario Gutiérrez:** Conceptualization, Methodology, Experiments, Data analysis and curation, Writing. **Cristina Martín:** Experiments, Data analysis, Writing. **Johan Hofkens**: Funding acquisition, Resources, Manuscript review. **Jin-Chong Tan:** Conceptualization, Methodology, Funding acquisition, Supervision, Resources, Writing.

## Conflicts of interest

There are no conflicts to declare.

## Acknowledgements

M.G. and J.C.T. thank the EPSRC IAA award (EP/R511742/1) and the ERC Consolidator Grant PROMOFS (grant agreement 771575) for funding the research. J.H. and C.M. acknowledge financial support from the Research Foundation - Flanders (FWO Grant Numbers S002019N, 1514220N, G.0B39.15, G.0B49.15, G098319N, and ZW15_09-GOH6316, the KU Leuven Research Fund (C14/19/079 and iBOF-21-085 PERSIST), the Flemish government through long term structural funding Methusalem (CASAS2, Meth/15/04). C.M. thanks the FWO for the fellowships received (12J1719N and 12J1716N). We thank Dr. C. Besnard and Prof. A.M. Korsunsky for access to the Scanning Electron Microscope facilities in the MBLEM Laboratory at Oxford.

## Notes and references

1.  A. Löffler; U. Tober, *Laser Tech. J.,* 2017, **14** (4), 19-21.
2.  S. Kasap; P. Capper, *Springer Handbook of Electronic and Photonic Materials.* Springer, Boston, MA: US, 2007.
3.  N. I. Zheludev, *Science,* 2015, **348** (6238), 973-974.
4.  Y. Hong; J. W. Y. Lam; B. Z. Tang, *Chem. Soc. Rev.,* 2011, **40** (11), 5361-5388.
5.  J. Mei; N. L. C. Leung; R. T. K. Kwok; J. W. Y. Lam; B. Z. Tang, *Chem. Rev.,* 2015, **115** (21), 11718-11940.
6.  Z. Zhao; H. Zhang; J. W. Y. Lam; B. Z. Tang, *Angew. Chem. Int. Ed.,* 2020, **59** (25), 9888-9907.
7.  X. Gan; G. Liu; M. Chu; W. Xi; Z. Ren; X. Zhang; Y. Tian; H. Zhou, *Org. Biomol. Chem.,* 2017, **15** (1), 256-264.
8.  K. Leduskrasts; E. Suna, *RSC Adv.,* 2019, **9** (1), 460-465.
9.  W. P. Lustig; S. Mukherjee; N. D. Rudd; A. V. Desai; J. Li; S. K. Ghosh, *Chem. Soc. Rev.,* 2017, **46** (11), 3242-3285.
10. Y. Cui; J. Zhang; H. He; G. Qian, *Chem. Soc. Rev.,* 2018, **47** (15), 5740-5785.
11. C. R. Martin; P. Kittikhunnatham; G. A. Leith; A. A. Berseneva; K. C. Park; A. B. Greytak; N. B. Shustova, *Nano Res.,* 2021, **14** (2), 338-354.
12. M. D. Allendorf; C. A. Bauer; R. K. Bhakta; R. J. T. Houk, *Chem. Soc. Rev.,* 2009, **38** (5), 1330-1352.
13. B. Chen; G. Qian, *Metal-Organic Frameworks for Photonics Applications.* Springer-Verlag Berlin Heidelberg: 2014.
14. H.-Q. Yin; X.-B. Yin, *Acc. Chem. Res.,* 2020, **53** (2), 485-495.
15. H.-R. Wang; X.-G. Yang; J.-H. Qin; L.-F. Ma, *Inorg. Chem. Front.,* 2021, **8** (8), 1942-1950.
16. R. Seetharaj; P. V. Vandana; P. Arya; S. Mathew, *Arab. J. Chem.,* 2019, **12** (3), 295-315.
17. A. D. Burrows; K. Cassar; R. M. W. Friend; M. F. Mahon; S. P. Rigby; J. E. Warren, *CrystEngComm,* 2005, **7** (89), 548-550.
18. B. Shan; J. B. James; M. R. Armstrong; E. C. Close; P. A. Letham; K. Nikkhah; Y. S. Lin; B. Mu, *J. Phys. Chem. C,* 2018, **122** (4), 2200-2206.
19. J. Hwang; R. Yan; M. Oschatz; B. V. K. J. Schmidt, *J. Mater. Chem. A,* 2018, **6** (46), 23521-23530.
20. H. Ma; A. Lv; L. Fu; S. Wang; Z. An; H. Shi; W. Huang, *Ann. Phys.,* 2019, **531** (7), 1800482.
21. W. Zhao; Z. He; B. Z. Tang, *Nat. Rev. Mater.,* 2020, **5** (12), 869-885.
22. A. Forni; E. Lucenti; C. Botta; E. Cariati, *J. Mater. Chem. C,* 2018, **6** (17), 4603-4626.
23. L.-L. Zhu; Y.-E. Huang; L.-K. Gong; X.-Y. Huang; X.-H. Qi; X.-H. Wu; K.-Z. Du, *Chem. Mater.,* 2020, **32** (4), 1454-1460.
24. Y. Fan; Q. Li; Z. Li, *Mater. Chem. Front.,* 2021, **5** (4), 1525-1540.
25. J. Yang; Z. Ren; B. Chen; M. Fang; Z. Zhao; B. Z. Tang; Q. Peng; Z. Li, *J. Mater. Chem. C,* 2017, **5** (36), 9242-9246.
26. M. Y. Wong; E. Zysman-Colman, *Adv. Mater.,* 2017, **29** (22), 1605444.
27. B. Ruan; H.-L. Liu; X.-Q. Zhan; H. Ding; L. Xie; F.-C. J. M. W. C. Tsai, *MATEC Web of Conferences,* 2018, **238**, 05004.
28. M. Ji; X. Lan; Z. Han; C. Hao; J. Qiu, *Inorg. Chem.,* 2012, **51** (22), 12389-12394.
29. F. G. Cirujano; F. X. Llabrés i Xamena, *J. Phys. Chem. Lett.,* 2020, **11** (12), 4879-4890.
30. S. Dissegna; K. Epp; W. R. Heinz; G. Kieslich; R. A. Fischer, *Adv. Mater.,* 2018, **30** (37), 1704501.
31. R. A. Dodson; A. P. Kalenak; A. J. Matzger, *J. Am. Chem. Soc.,* 2020, **142** (49), 20806-20813.
32. V. M. Saasa, T.; Beukes, M.; Mokgotho, M.; Liu, C.-P.; Mwakikunga, B., *Diagnostics,* 2018, **8**, 12.
33. A. Rydosz, *Sensors,* 2018, **18**, 2298.
34. A. S. Amann, D. , *Volatile Biomarkers. In Non-Invasive Diagnosis in Physiology and Medicine.* Elsevier: New York, NY, USA,: 2013.
35. D. Manzani; J. F. d. S. Petruci; K. Nigoghossian; A. A. Cardoso; S. J. L. Ribeiro, *Sci. Rep.,* 2017, **7**, 41596.
36. W. Xu; H. Zhao; Z. Zhang; W. Cao, *Sens. Actuators, B,* 2013, **178**, 520-524.
37. J. Rocha; C. D. S. Brites; L. D. Carlos, *Chem. Eur. J.,* 2016, **22** (42), 14782-14795.
38. Y. Cui; F. Zhu; B. Chen; G. Qian, *Chem. Commun.,* 2015, **51** (35), 7420-7431.
39. A. M. P. Peedikakkal; M. Qamar, *J. Chem. Sci.,* 2018, **130** (5), 45.
40. M. Gutiérrez; C. Martín; K. Kennes; J. Hofkens; M. Van der Auweraer; F. Sánchez; A. Douhal, *Adv. Opt. Mater,* 2018, **6** (6), 1701060.
41. M. Gutiérrez; C. Martín; M. Van der Auweraer; J. Hofkens; J.-C. Tan, *Adv. Opt. Mater,* 2020, **8** (16), 2000670.
42. M. Gutiérrez; C. Martín; B. E. Souza; M. Van der Auweraer; J. Hofkens; J.-C. Tan, *Appl. Mater. Today,* 2020, **21**, 100817.
43. X.-G. Yang; X.-M. Lu; Z.-M. Zhai; Y. Zhao; X.-Y. Liu; L.-F. Ma; S.-Q. Zang, *Chem. Commun.,* 2019, **55** (74), 11099-11102.

44. Z. Wang; C.-Y. Zhu; P.-Y. Fu; J.-T. Mo; J. Ruan; M. Pan; C.-Y. Su, *Chem. Eur. J.,* 2020, **26** (33), 7458-7462.

45. Q. Li; Z. Li, *Acc. Chem. Res.,* 2020, **53** (4), 962-973.

46. M.-L. Zhang; Y. Bai; X.-G. Yang; Y.-J. Zheng; Y.-X. Ren; J.-J. Wang; M.-L. Han; F.-F. Li; L.-F. Ma, *Dalton Trans.,* 2020, **49** (29), 9961-9964.

47. Y. Zhao; X.-G. Yang; X.-M. Lu; C.-D. Yang; N.-N. Fan; Z.-T. Yang; L.-Y. Wang; L.-F. Ma, *Inorg. Chem.,* 2019, **58** (9), 6215-6221.